\documentclass[review,12pt]{elsarticle}

\usepackage{amssymb}
\usepackage{amsthm,amsmath}
\usepackage{accents}
\usepackage{graphicx}
\usepackage{epstopdf}
\newtheorem{theorem}{Theorem}

\newtheorem{col}{Corollary}
\newtheorem{remark}{Remark}

\begin{document}
\begin{center}
{\Large  New approach to quasi synchronization of coupled heterogeneous complex networks }

\footnote{This work is jointly supported by the National Key R$\&$D Program of China (No.
2018AAA010030), National Natural Sciences Foundation of China under Grant (No.
61673119 and 61673298).
}
\\[0.2in]
\begin{center}
Tianping Chen\footnote{Tianping Chen is with the School of Computer
Sciences/Mathematics, Fudan University, 200433, Shanghai, China. \\
 Email:
tchen@fudan.edu.cn},

\end{center}
\end{center}

\section{abstract}
This short paper addresses quasi synchronization
of heterogeneous complex dynamical networks. The similarity and difference between synchronization of homogeneous systems and quasi synchronization
of heterogeneous complex dynamical networks will be revealed.

Key words Quasi synchronization, complex network, heterogeneous,
homogeneous, synchronization.



\section{Introduction}
Synchronization of complex networks has been a hot issue for decades.

In [1], a general framework is presented to analyze
synchronization stability of Linearly Coupled Ordinary
Differential Equations (LCODEs). The uncoupled dynamical behavior
at each node is general, which can be chaotic or others; the
coupling configuration is also general, without assuming the
coupling matrix be symmetric or irreducible.

It is clear that complete synchronization applies only to homogeneous systems. Later, researchers discussed synchronization of heterogeneous systems or non-identical nodes. For example, in [2], the cluster synchronization of networks with nonidentical nodes was addressed. In [3]-[12], the quasi synchronization or bounded synchronization was investigated.

Based on the approach and techniques proposed in [1], in this paper, we proposed a new approach to discuss quasi-synchronization of complex networks.

This article is organized as follows. Section 3 introduces
the network model and some mathematical preliminaries. Section 4
presents our main results. Section 5 concludes the article.

\section{Models and basic mathematical concepts}

In [1], the synchronization of the coupled homogeneous systems
\begin{align}\label{modelhomo}
\frac{d x_{i}(t)}{dt}=f(x_{i}(t))
+c\sum\limits_{j=1}^{m}a_{ij} x_{j}(t),\quad i=1,\cdots,m
\end{align}
were discussed. Here, every node $x_{i}(t)=[x_{i}^{1}(t),\cdots,x_{i}^{1}(t)]^{T}\in R^{n}$, $i=1,\cdots,m$, has same intrinsic dynamics $\dot{x}_{i}(t)=f(x_{i}(t))$.

Synchronization of the system (1) is defined as
\begin{align}\label{def1}
\lim_{t\rightarrow\infty}||x_{i}(t)-x_{j}(t)||=0,\quad i=1,\cdots,m
\end{align}

Instead, in this paper, we discuss quasi synchronization of the following heterogeneous systems
\begin{align}\label{modelhetero}
\frac{d x_{i}(t)}{dt}=f_{i}(x_{i}(t))
+c\sum\limits_{j=1}^{m}a_{ij} x_{j}(t),\quad i=1,\cdots,m
\end{align}
where every node has different intrinsic dynamics $\dot{x}_{i}(t)=f_{i}(x_{i}(t))$.

Quasi synchronization of the system (3) is defined as
\begin{align}\label{def1}
\lim_{t\rightarrow\infty}||x_{i}(t)-x_{j}(t)||\le \delta,\quad i=1,\cdots,m
\end{align}
for some positive constant $\delta>0$.

Readers can also refer to [3]-[12].

Throughout the paper, we assume that the coupling matrix $A=(a_{ij})$ is a connected  Mezler matrix, i.e. $a_{ij}\ge 0$, if $i\ne j$ and $a_{ii}=-\sum_{j\ne i}a_{ij}$. $Rank(A)=m-1$.

Let~$\Xi=[\xi_{1},\cdots,\xi_{m}]^{T}$~be the left eigenvector corresponding to eigenvalue $0$ of the coupling matrix $A$.

It was proved in [1] that the eigenvalues of $\Xi A+A^{T}\Xi$ can be written as $0=\lambda_{1}>\lambda_{2}\ge \cdots\ge \lambda_{m}$.

For any trajectory $x(t)=[x_{1}^{T}(t),\cdots,x_{m}^{T}(t)]^{T}$, denote $\bar{x}(t)=[\bar{x}^{1}(t),\cdots,\bar{x}^{n}(t)]^{T}=\sum_{i=1}^{m}\xi_{i}x_{i}(t)$. $\bar{X}(t)=[\bar{x}^{T}(t),\cdots,\bar{x}^{T}(t)]^{T}$.

It was reported in [1]
\begin{align}
\sum_{i=1}^{m}\xi_{i}a_{ij}=0,~ \sum_{i=1}^{m}\xi_{i}(x_{i}(t)-\bar{x}(t))J(t)=0
\end{align}
where $J(t)$ is any function independent of the index $i$. Thus,
\begin{align}
\sum_{i=1}^{m}\xi_{i}(x_{i}(t)-\bar{x}(t))\dot{\bar{x}}(t)=0
\end{align}

\section{Main results}

Define two norms $||x(t)-\bar{X}(t)||$ and $||x(t)-\bar{X}(t)||_{\xi}$ by
\begin{align}
||x(t)-\bar{X}(t)||^{2}=
\sum_{i=1}^{m}[x_{i}(t)-\bar{x}(t)]^{T}[x_{i}(t)-\bar{x}(t)]
\end{align}
\begin{align}
||x(t)-\bar{X}(t)||_{\xi}^{2}=
\sum_{i=1}^{m}\xi_{i}[x_{i}(t)-\bar{x}(t)]^{T}[x_{i}(t)-\bar{x}(t)]
\end{align}

It is clear that
\begin{align}
\underline{\xi}||x(t)-\bar{X}(t)||^{2}\le||x(t)-\bar{X}(t)||_{\xi}^{2}\le \bar{\xi}||x(t)-\bar{X}(t)||^{2}
\end{align}
where
$$\underline{\xi}=\min_{i=1,\cdots,m}\xi_{i},~ \bar{\xi}=\max_{i=1,\cdots,m}\xi_{i}$$

Differentiating $||x(t)-\bar{X}(t)||_{\xi}^{2}$, we have
\begin{align}
\frac{d}{dt}||x(t)-\bar{X}(t)||_{\xi}^{2}&
=\sum_{i=1}^{m}\xi_{i}[x_{i}(t)-\bar{x}(t)]^{T}[f_{i}(x_i(t)-J(t)]\nonumber\\&
+c\sum_{i=1}^{m}\sum_{j=1}^{m}(x_i(t)-\bar{x}(t))^{T}[\xi_{i} a_{ij}+\xi_{i} a_{ij}]
(x_j(t)-\bar{x}(t))\nonumber\\&=\tilde{K}_1(t)+\tilde{K}_2(t)
\end{align}
where $J(t)$ can be any function independent of index $i$. Here, we can choose $J(t)=\sum_{i=1}^{m}\xi_{i}f_{i}(x_{i}(t))$,

It can be seen that $\tilde{K}_{2}(t)$ describes the effect of the coupled term, which helps synchronization. Instead, $\tilde{K}_{1}(t)$ describes the effect of deviation between nodes, which should be overcome.

Simple calculations show
\begin{align}
\tilde{K}_2(t)
&\le c\lambda_{2}\sum_{j=1}^{m}(x_i(t)-\bar{x}(t))^{T}(x_j(t)-\bar{x}(t))
\nonumber\\
&\le \frac{c\lambda_{2}}{\overline{\xi}}\sum_{j=1}^{m}\xi_{i}(x_i(t)-\bar{x}(t))^{T}(x_j(t)-\bar{x}(t))
\nonumber\\
&=\frac{c\lambda_{2}}{\overline{\xi}}||x(t)-\bar{X}(t)||_{\xi}^{2}
\end{align}

Notice $\sum_{j=1}^{m}\xi_{j}=1$, by Jessen inequality
$$(\sum_{i=1}^{m}\xi_{i}a_{i})^{1/2}\ge (\sum_{i=1}^{m}\xi_{i}a_{i}^{1/2})$$
we have
\begin{align}
&|\sum_{i=1}^{m}\xi_{i}[x_{i}(t)-\bar{x}(t)]^{T}[f_{i}(x_i(t)-J(t)]|\le \tilde{c}(t)|\sum_{i=1}^{m}\sum_{k=1}^{n}\xi_{i}|x_{i}^{k}(t)-\bar{x}^{k}(t)|\nonumber
\\&\le
\tilde{c}(t)(\sum_{i=1}^{m}\xi_{i}[x_{i}(t)-\bar{x}(t)]^{T}[x_{i}(t)-\bar{x}(t)])^{1/2}
=\tilde{c}(t)||x(t)-\bar{X}(t)||_{\xi}
\end{align}
where
\begin{align}
\tilde{c}(t)=\sup_{i=1,\cdots,m}||f_{i}(x_i(t)-J(t)||
\end{align}

Therefore,
\begin{align}\label{K1a}
\tilde{K}_1(t)=&\sum_{i=1}^{m}\xi_{i}[x_{i}(t)-\bar{x}(t)]^{T}[f(x_i(t)-J(t)]\nonumber\\
&\le \tilde{c}(t)||x(t)-\bar{X}(t)||_{\xi}
\end{align}

Combining them, we obtain
\begin{align}
\frac{d}{dt}||x(t)-\bar{X}(t)||_{\xi}^{2}
<-\frac{c|\lambda_{2}|}{\overline{\xi}}||x(t)-\bar{X}(t)||_{\xi}^{2}
+\tilde{c}(t)||x(t)-\bar{X}(t)||_{\xi}
\end{align}

In case that
$$||x(t)-\bar{X}(t)||_{\xi}
>\frac{\tilde{c}(t)\overline{\xi}}{c|\lambda_{2}|}$$
we have
\begin{align}
\frac{d}{dt}||x(t)-\bar{X}(t)||_{\xi}^{2}
<0
\end{align}
and
\begin{align}
\lim_{t\rightarrow\infty}||x(t)-\bar{X}(t)||_{\xi}\le \frac{\overline{\xi}}{c|\lambda_{2}|}\overline{lim}_{t\rightarrow\infty}\tilde{c}(t)
\end{align}
which implies
\begin{align}
\overline{lim}_{t\rightarrow\infty}||x(t)-\bar{X}(t)||\le \frac{\overline{\xi}}{c|\lambda_{2}|\sqrt{\underline{\xi}}}
\overline{lim}_{t\rightarrow\infty}\tilde{c}(t)
\end{align}

In summary, we have

\begin{theorem}
Suppose that there exists a function $J(t)$ and let $$\tilde{c}(t)=sup_{\{i=1,\cdots,m\}}||f_{i}(x_i(t)-J(t)||$$
Then, the trajectory of the system  (3) satisfies
\begin{align}
\overline{lim}_{t\rightarrow\infty}||x(t)-\bar{x}(t)||\le \frac{\overline{\xi}}{c|\lambda_{2}|\sqrt{\underline{\xi}}}
\overline{lim}_{t\rightarrow\infty}\tilde{c}(t)
\end{align}
i.e. the system  (3) reaches quasi synchronization.
\end{theorem}

\begin{remark}
The constant $c$ is the coupling strength, $\lambda_{2},\overline{\xi},\underline{\xi}$ are decided by the coupling matrix $A$. The function $\tilde{c}(t)$ plays key role, which depends heavily upon the $f_{i}(x)$ and selection of $J(t)$. The smaller $\tilde{c}(t)$ is, the better result of synchronization is.
\end{remark}

It is well known that if $A$ is symmetric, then all $\xi_{i}=\frac{1}{m}$. Therefore, we have
\begin{col}
Suppose that the coupling matrix $A$ is symmetric and there exists a function $J(t)$ and let $$\tilde{c}(t)=sup_{\{i=1,\cdots,m\}}||f_{i}(x_i(t))-J(t)||$$
Then, the trajectory of the system (3) satisfies
\begin{align}
\overline{lim}_{t\rightarrow\infty}||x(t)-\bar{x}(t)||\le \frac{1}{c|\lambda_{2}|\sqrt{m}}
\overline{lim}_{t\rightarrow\infty}\tilde{c}(t)
\end{align}
i.e. the system  (3) reaches quasi synchronization.
\end{col}

\begin{remark}
In [1], the complete synchronization of the coupled homogeneous systems
\begin{align}
\frac{d x_{i}(t)}{dt}=f(x_{i}(t))
+c\sum\limits_{j=1}^{m}a_{ij} x_{j}(t),\quad i=1,\cdots,m
\end{align}
was discussed. Here, every node has same intrinsic dynamics $\dot{x}_{i}(t)=f(x_{i}(t))$.

In this case,
we have
\begin{align}
\frac{d}{dt}||x(t)-\bar{X}(t)||_{\xi}^{2}&
=\sum_{i=1}^{m}\xi_{i}[x_{i}(t)-\bar{x}(t)]^{T}[f(x_i(t)-f(\bar{x}(t)]\nonumber\\&
+c\sum_{i=1}^{m}\sum_{j=1}^{m}(x_i(t)-\bar{x}(t))[\xi_{i} a_{ij}+\xi_{i} a_{ij}]
(x_j(t)-\bar{x}(t))\nonumber\\&=K_1(t)+K_2(t)
\end{align}
where 
\begin{align}
K_2(t)
\le\frac{c\lambda_{2}}{\overline{\xi}}||x(t)-\bar{X}(t)||_{\xi}^{2}
\end{align}
\begin{align}K_{1}(t)&=\sum_{i=1}^{m}\xi_{i}[x_{i}(t)-\bar{x}(t)]^{T}[f(x_i(t)-f(\bar{x}(t)]
\nonumber\\
&\le c(t)\sum_{i=1}^{m}\xi_{i}[x_{i}(t)-\bar{x}(t)]^{T}[x_{i}(t)-\bar{x}(t)]\nonumber\\
&=c(t)||x(t)-\bar{X}(t)||_{\xi}^{2}
\end{align}
which implies
\begin{align}
\frac{d}{dt}||x(t)-\bar{X}(t)||_{\xi}^{2}
<-\frac{c|\lambda_{2}|}{\overline{\xi}}||x(t)-\bar{X}(t)||_{\xi}^{2}
+c(t)||x(t)-\bar{X}(t)||_{\xi}^{2}<0
\end{align}
Therefore, if $\frac{c|\lambda_{2}|}{\overline{\xi}}>c(t)$, synchronization is reached.

\end{remark}
\begin{remark}
For homogeneous system, $J(t)=f(\bar{x}(t))$ and
\begin{align}
\frac{d}{dt}||x(t)-\bar{X}(t)||_{\xi}^{2}
<-\frac{c|\lambda_{2}|}{\overline{\xi}}||x(t)-\bar{X}(t)||_{\xi}^{2}
+c(t)||x(t)-\bar{X}(t)||_{\xi}^{2}<0
\end{align}
$$J(t)=f(\bar{x}(t)),~~K_{1}(t)\le c(t)||x(t)-\bar{X}(t)||_{\xi}^{2}$$

Instead, for heterogeneous systems, $J(t)=\sum_{i=1}^{m}\xi_{i}f_{i}(\bar{x}(t))$ and
\begin{align}
\frac{d}{dt}||x(t)-\bar{X}(t)||_{\xi}^{2}
<-\frac{c|\lambda_{2}|}{\overline{\xi}}||x(t)-\bar{X}(t)||_{\xi}^{2}
+\tilde{c}(t)||x(t)-\bar{X}(t)||_{\xi}
\end{align}
\end{remark}

\section{Conclusions}
In this short paper, quasi synchronization
of heterogeneous complex dynamical networks is discussed. An effective theoretical analysis 
is provided.

The dynamics of the coupled homogeneous complex networks is revealed. The comparisons between 
dynamics for homogeneous complex networks and heterogeneous complex networks are revealed.

\end{document}